\documentclass{emulateapj}

\newcommand{\psr}{PSR~J1024$-$0719}
\newcommand{\masyr}{\ensuremath{{\rm mas}\,{\rm yr}^{-1}}}
\newcommand{\expnt}[2]{\ensuremath{#1 \times 10^{#2}}}   
\newcommand{\kms}{\ensuremath{{\rm km}\,{\rm s}^{-1}}}

\newcommand{\bright}{1024-Br}
\newcommand{\faint}{1024-Fnt}

\slugcomment{Accepted to ApJ}

\begin{document}

\title{\psr: A Millisecond Pulsar in an Unusual Long-Period Orbit}
\author{%
David~L.~Kaplan$^{1}$,
Thomas~Kupfer$^{2}$,
David~J.~Nice$^{3}$,
Andreas~Irrgang$^{4}$,
Ulrich~Heber$^{4}$,
Zaven~Arzoumanian$^{5}$,
Elif~Beklen$^{6,7}$,
Kathryn~Crowter$^{8}$,
Megan~E.~DeCesar$^{1}$,
Paul~B.~Demorest$^{9}$,
Timothy~Dolch$^{10}$,
Justin~A.~Ellis$^{11}$,
Robert~D.~Ferdman$^{12}$,
Elizabeth~C.~Ferrara$^{13}$,
Emmanuel~Fonseca$^{8}$,
Peter~A.~Gentile$^{7}$,
Glenn~Jones$^{14}$,
Megan~L.~Jones$^{7}$,
Simon~Kreuzer$^{4}$,
Michael~T.~Lam$^{15}$,
Lina~Levin$^{7}$,
Duncan~R.~Lorimer$^{7}$,
Ryan~S.~Lynch$^{16}$,
Maura~A.~McLaughlin$^{7}$,
Adam~A.~Miller$^{2,11,17}$,
Cherry~Ng$^{8}$,
Timothy~T.~Pennucci$^{18}$,
Tom~A.~Prince$^{2}$,
Scott~M.~Ransom$^{19}$,
Paul~S.~Ray$^{20}$,
Renee~Spiewak$^{1}$,
Ingrid~H.~Stairs$^{8,21}$,
Kevin~Stovall$^{22}$,
Joseph~Swiggum$^{1}$
and
Weiwei~Zhu$^{23}$
}

\address{$^{1}$~Center for Gravitation, Cosmology and Astrophysics, Department of Physics, University of Wisconsin-Milwaukee,\\ P.O. Box 413, Milwaukee, WI 53201, USA}
\address{$^{2}$~Division of Physics, Mathematics, and Astronomy, California Institute of Technology, Pasadena, CA 91125, USA}
\address{$^{3}$~Department of Physics, Lafayette College, Easton, PA 18042, USA}
\address{$^{4}$~Dr. Karl Remeis-Observatory \& ECAP, Astronomical Institute, Friedrich-Alexander University Erlangen-Nuremberg, Sternwartstr. 7, 96049 Bamberg, Germany}
\address{$^{5}$~Center for Research and Exploration in Space Science and Technology and X-Ray Astrophysics Laboratory, \\NASA Goddard Space Flight Center, Code 662, Greenbelt, MD 20771, USA}
\address{$^{6}$~Department of Physics, Suleyman Demirel University, 32260, Isparta, Turkey}
\address{$^{7}$~Department of Physics and Astronomy, West Virginia University, P.O. Box 6315, Morgantown, WV 26506, USA}
\address{$^{8}$~Department of Physics and Astronomy, University of British Columbia, 6224 Agricultural Road, Vancouver, BC V6T 1Z1, Canada}
\address{$^{9}$~National Radio Astronomy Observatory, 1003 Lopezville Rd., Socorro, NM 87801, USA}
\address{$^{10}$~Department of Physics, Hillsdale College, 33 E. College Street, Hillsdale, Michigan 49242, USA}
\address{$^{11}$~Jet Propulsion Laboratory, California Institute of Technology, 4800 Oak Grove Drive, Pasadena, CA 91109, USA}
\address{$^{12}$~Department of Physics, McGill University, 3600  University St., Montreal, QC H3A 2T8, Canada}
\address{$^{13}$~NASA Goddard Space Flight Center, Greenbelt, MD 20771, USA}
\address{$^{14}$~Department of Physics, Columbia University, New York, NY 10027, USA}
\address{$^{15}$~Department of Astronomy, Cornell University, Ithaca, NY 14853, USA}
\address{$^{16}$~National Radio Astronomy Observatory, P.O. Box 2, Green Bank, WV 24944, USA}
\address{$^{17}$~Hubble Fellow}
\address{$^{18}$~University of Virginia, Department of Astronomy, P.O. Box 400325, Charlottesville, VA 22904, USA}
\address{$^{19}$~National Radio Astronomy Observatory, 520 Edgemont Road, Charlottesville, VA 22903, USA}
\address{$^{20}$~Space Science Division, Naval Research Laboratory, Washington, DC 20375-5352, USA}
\address{$^{21}$~McGill Space Institute, 3550 Rue University, Montreal, Quebec H3A 2A7 Canada}
\address{$^{22}$~Department of Physics and Astronomy, University of New Mexico, Albuquerque, NM 87131, USA}
\address{$^{23}$~Max Planck Institute for Radio Astronomy, Auf dem H\"{u}gel 69, D-53121 Bonn, Germany}

\begin{abstract}
\psr\ is a millisecond pulsar that was long thought to be
isolated.  However, puzzling results concerning its velocity,
distance, and low rotational period derivative have led to reexamination of its
properties.  We present updated radio timing observations along with
new and archival optical data that show \psr\ is most likely in a long
period (2--20\,kyr) binary system with a low-mass ($\approx
0.4\,M_\odot$) low-metallicity ($Z \approx -0.9$\,dex) main
sequence star.  Such a system can explain most of the anomalous
properties of this pulsar.  We suggest that this system formed through
a dynamical exchange in a globular cluster that ejected it into a halo
orbit, consistent with the low observed metallicity for the stellar
companion.  Further astrometric and radio timing observations such as
measurement of the third period derivative could strongly constrain the
range of orbital parameters.
\end{abstract} 


\section{Introduction}\label{sec:intro}

\subsection{Pulsar characteristics and early distance estimates}\label{sec:context}

\object[PSR J1024-0719]{\psr}\ is a millisecond pulsar (MSP)  with rotation period $P=5.2$~ms and
period derivative $\dot{P}=1.8\times 10^{-20}$
\citep{1997ApJ...481..386B},  
typical of other MSPs.
There was no evidence
for binary motion in timing observations of the pulsar, so it was regarded as isolated.
Its dispersion measure,
${\rm DM}=6.5\,{\rm pc\,cm}^{-3}$, is among the lowest measured; it
implies a distance $d_{\rm DM}\approx 0.390$~kpc based on the \cite{cl02}
Galactic electron density model.

A second line of reasoning also led to similar distance estimates.
Observed pulsar period derivatives $\dot P_{\rm obs}$ are biased from their intrinsic
values $\dot P_{\rm int}$ according to the Shklovskii effect, $\dot{P}_{\rm obs} =
\dot{P}_{\rm int} + \dot{P}_{\rm Shk}$ with $\dot P_{\rm
  Shk}=\mu^2dP/c$, where $\mu$ is the proper motion, $d$ is the
  distance, and $c$ is the speed of light \citep{shklovskii70}.  For
  a pulsar losing rotational energy, the intrinsic spin-down rate
  must be positive, $\dot{P}_{\rm int}>0$, so the Shklovskii effect
  places an upper limit on distance, $d_{\dot{P}}<\dot{P}_{\rm
    obs}c/\mu^2P$.  \cite{1999MNRAS.307..925T} used an early proper
  motion measurement, $\mu\approx 81~\masyr$, to place an upper limit
  $d_{\dot{P}}<0.226$~kpc.  Later measurements revised the proper
  motion down to $\mu\approx 60$~mas which
  gives $d_{\dot{P}}<0.430$~kpc, consistent with $d_{\rm DM}$ \citep{2006MNRAS.369.1502H}.


\subsection{Optical and $\gamma$-ray observations}\label{sec:introoptical}

\citet{2003A&A...406..245S} searched for an optical
counterpart to \psr\ using deep observations with the Very Large
Telescope (VLT). They found two potential counterparts: a faint one
(\faint, with $R=24.4$) and a bright one, which appeared
to be a K star (\bright, with
$R=18.9$).  While they found the position of \bright\ to be 
coincident with that of the pulsar to within $0\farcs2$, they rejected
an association between \bright\ and the pulsar because (i) \bright\
is much more distant than the distance estimate for the pulsar 
then available;
(ii) the proper motion of \bright, which they
estimated by comparing their observations with earlier catalogs,
disagreed with the proper motion of the pulsar
reported by \citet{1999MNRAS.307..925T}; and (iii)
the very small timing residuals of the pulsar suggested
that it was an isolated object, with no evidence for binary motion.

\citet{2013MNRAS.430..571E} identified a \textit{Fermi} $\gamma$-ray
counterpart to \psr.  They assumed that the pulsar was at a distance
$d=0.4$~kpc, close to
the maximum allowed by the Shklovskii effect (they used
$d<0.410$~kpc).  Correcting for the Shklovskii effect, they estimated
the intrinsic period derivative to be 
$\dot{P}_{\rm int}\le 5\times 10^{-22}$, an unusually
small value, which implies an unusually small rotational
energy loss rate, $\dot{E}\propto\dot{P}/P^3$.
This, in turn, yielded a value for $\gamma$-ray
efficiency,
$\eta=L_\gamma/\dot E>0.8$,  higher than typically found.


\subsection{Recent developments}\label{sec:recent}

\psr\ is under observation by several groups as part of the global effort to
detect nanohertz gravitational waves via millisecond pulsar timing.

Four recent papers used high precision timing to measure or constrain
its parallax, and hence its distance; we summarize these measurements in
Table~\ref{tab:parallax}.  
All of these recent pulsar timing distances are consistent
with each other, and they are incompatible with the distance upper
limit $d_{\dot{P}}< 0.430$~kpc derived from the Shklovskii effect using
the latest proper motion measurements
\citep[e.g.,][]{2016ApJ...818...92M}.  Three of these papers argued
that the discrepancy between these distances could be resolved if the
pulsar were undergoing acceleration, $\dot{v}$, due to gravitational
interaction with another object
\citep{2016ApJ...818...92M,2016arXiv160105987G,2016arXiv160208511D}.
This would add a further bias to the observed spin-down rate,
$\dot{P}_{\rm obs} = \dot{P}_{\rm int} + \dot{P}_{\rm Shk} + \dot{P}_{\rm orb}$,
where $\dot{P}_{\rm orb}={\dot v}P/c$.  If the pulsar were undergoing
negative acceleration, $\dot v<0$, the
resulting negative $\dot{P}_{\rm orb}$ term would allow for a larger positive
$\dot{P}_{\rm Shk}=\mu^2d/c$ term, which in turn would allow it to
contain a larger distance than the previous upper limit  of 0.430~kpc.


Such a gravitational interaction (i.e., an orbit) could potentially be
manifested in timing observations as a rotation period second
derivative, $\ddot{P}$, due to the jerk, or change in acceleration, of
the pulsar (details discussed below).  \citet{2016ApJ...818...92M}
combined their data with previous observations and placed an upper
limit $\ddot{P}\lesssim 1\times 10^{-23}~{\rm yr}^{-1} = 3\times
10^{-31}~{\rm s}^{-1}$.  Considering this and other constraints, they
found that the pulsar acceleration could be caused by an orbit with
period greater than about 14,000~years and a companion star mass of
0.1~M$_{\odot}$ or greater, and they noted that \bright\ satisfied this
mass constraint.

\cite{2016arXiv160105987G} measured $\ddot{P}
=(2.2\pm0.2)\times 10^{-24}~{\rm yr}^{-1}
=(7.0\pm 0.6)\times 10^{-32}~{\rm s}^{-1}$, a
value just under the limit of \citet{2016ApJ...818...92M}.
\cite{2016MNRAS.455.1751R} and \cite{2016arXiv160208511D} reported red
timing noise (see also \citealt{2015arXiv151009194C}); such red noise
could be indicative of a nonzero value of $\ddot{P}$ not accounted for
in the timing model applied to their data.

\begin{deluxetable*}{l c c l}
  \tablewidth{0pt}
\tablecaption{Parallax and Distance Measurements for \psr\label{tab:parallax}}
\tablehead{
\colhead{Type} &  \colhead{Parallax} & \colhead{Distance} &
\colhead{Reference} \\
 & \colhead{(mas)} & \colhead{(kpc)} & 
}
\startdata
Dispersion Measure & \nodata & 0.39 & \citealt{cl02} \\
$\dot P$ limit & \nodata & $<0.43$ & \citealt{2006MNRAS.369.1502H}\\
NANOGrav\tablenotemark{a} 9-year Timing & $<1.10$ & $>0.91$ & \citealt{2016ApJ...818...92M}\\
PPTA\tablenotemark{b} Timing & $0.5\pm0.3$ & $1.1_{-0.3}^{+0.4}$\tablenotemark{d} &
\citealt{2016MNRAS.455.1751R} \\
Nan\c{c}ay Timing & $0.89\pm0.14$ & $1.13\pm0.18$ &
\citealt{2016arXiv160105987G} \\
EPTA\tablenotemark{c} Timing & $0.80\pm0.17$ & $1.08_{-0.16}^{+0.23}$\tablenotemark{d} &
\citealt{2016arXiv160208511D} \\
NANOGrav\tablenotemark{a} 11-year Timing & $0.77\pm0.23$ & $1.3_{-0.3}^{+0.6}$ & this paper\\
1024-Br Spectrum Main-sequence Fit\tablenotemark{e} & \nodata & $1.08\pm0.04$ & this paper
\enddata
\tablenotetext{a}{North American Nanohertz Observatory for Gravitational
  waves, \url{http://www.nanograv.org}.}
\tablenotetext{b}{Parkes
Pulsar Timing Array, \url{http://www.atnf.csiro.au/research/pulsar/ppta/}.}
\tablenotetext{c}{European Pulsar Timing Array,
\url{http://www.epta.eu.org}.}
\tablenotetext{d}{Includes adjustment for Lutz-Kelker bias
  \citep{2012ApJ...755...39V}.}
\tablenotetext{e}{For an assumed companion mass of $0.4\,M_\odot$.}
\end{deluxetable*}

\subsection{This paper}

In this paper, we argue that \psr\ is in a long-period (2--20\,kyr)
binary system with \bright.  In \S\ref{sec:timing} we present updated
NANOGrav timing observations of \psr, including new parallax and
period second derivative measurements.  In \S\ref{sec:optical}, we
present astrometric analysis from new and archival optical data for
\bright\ and show that its position and proper motion are completely
consistent with those of the pulsar, leaving no doubt that they are a
common proper motion pair.  Additionally we present a spectroscopic
analysis of \bright\ and show that the companion is consistent with a
star of spectral type K or M. In \S\ref{sec:binary}, we use constraints on
the position offset, acceleration, and jerk in this system to analyze
possible binary system parameters, for both circular and generalized
orbits.  We find the binary to be very wide, and the pulsar space
velocity to be unusually fast.  In \S\ref{sec:discuss} we discuss
formation scenarios for such a system.  In \S\ref{sec:conclusions} we
summarize our results.    

Unless otherwise noted, proper motions in
right ascension $\alpha$ are $\mu_\alpha=\dot \alpha \cos \delta$ in
units of mas\,yr$^{-1}$, and all positions are J2000.

During the preparation of this paper, we
became aware that another group had come to similar conclusions
regarding the nature of \psr\ \citep{bjs+16}.  Our analysis is very
similar to that presented in \citet{bjs+16}, although our data (aside
from archival optical observations) are entirely independent.  
\citet{bjs+16} additionally present an alternate formation mechanism for
\psr\ which we discuss briefly in \S\ref{sec:discuss}.

\section{Radio Timing}
\label{sec:timing}
We made radio timing observations of \psr\ over 6.3 years using the
100-m Robert C.\ Byrd Green Bank Telescope.
The resulting data will be part of
the upcoming NANOGrav 11-year data set (Arzoumanian et al 2016, in
preparation).  The observation and data-reduction procedures are
nearly identical to those of the NANOGrav 9-year data set 
\citep{2015ApJ...813...65T}.  Briefly, pulse arrival times were
made using two separate receiver systems, near 820 and 1400\,MHz, at
roughly monthly intervals.  The arrival times were fit to a standard
timing model using the {\sc tempo}
package\footnote{\url{http://tempo.sourceforge.net}}.  The timing
model included: astrometric parameters; independent dispersion measure
at every epoch (where epoch is defined as a period of six days or less); 
a white noise model; and a pulsar frequency model as
described below.  The JPL DE430 Solar System ephemeris \citep{2014IPNPR.196C...1F}
was used for Earth motion around the solar system, so astrometric values
are relative to this frame, which in turn is tied to the Second Realization
of the International Celestial
Reference Frame (ICRF2).  
The ephemeris was rotated by 
$23^\circ 26\arcmin 21\farcs 406$, the IERS2010 obliquity of the ecliptic,
to give position and proper motion in ecliptic coordinates.
Arrival times were adjusted following the TT(BIPM15)\footnote{\url{ftp://tai.bipm.org/TFG/TT(BIPM)}.} time scale, and parameters are ultimately
presented in Barycentric Dynamical Time (TDB).
Besides using updated ephemeris and time standards,
the primary difference between our work and the analysis procedure of
\cite{2015ApJ...813...65T}
is our use of frequency derivatives, as described below, instead of
a red noise model. 

\begin{figure*}
  \plotone{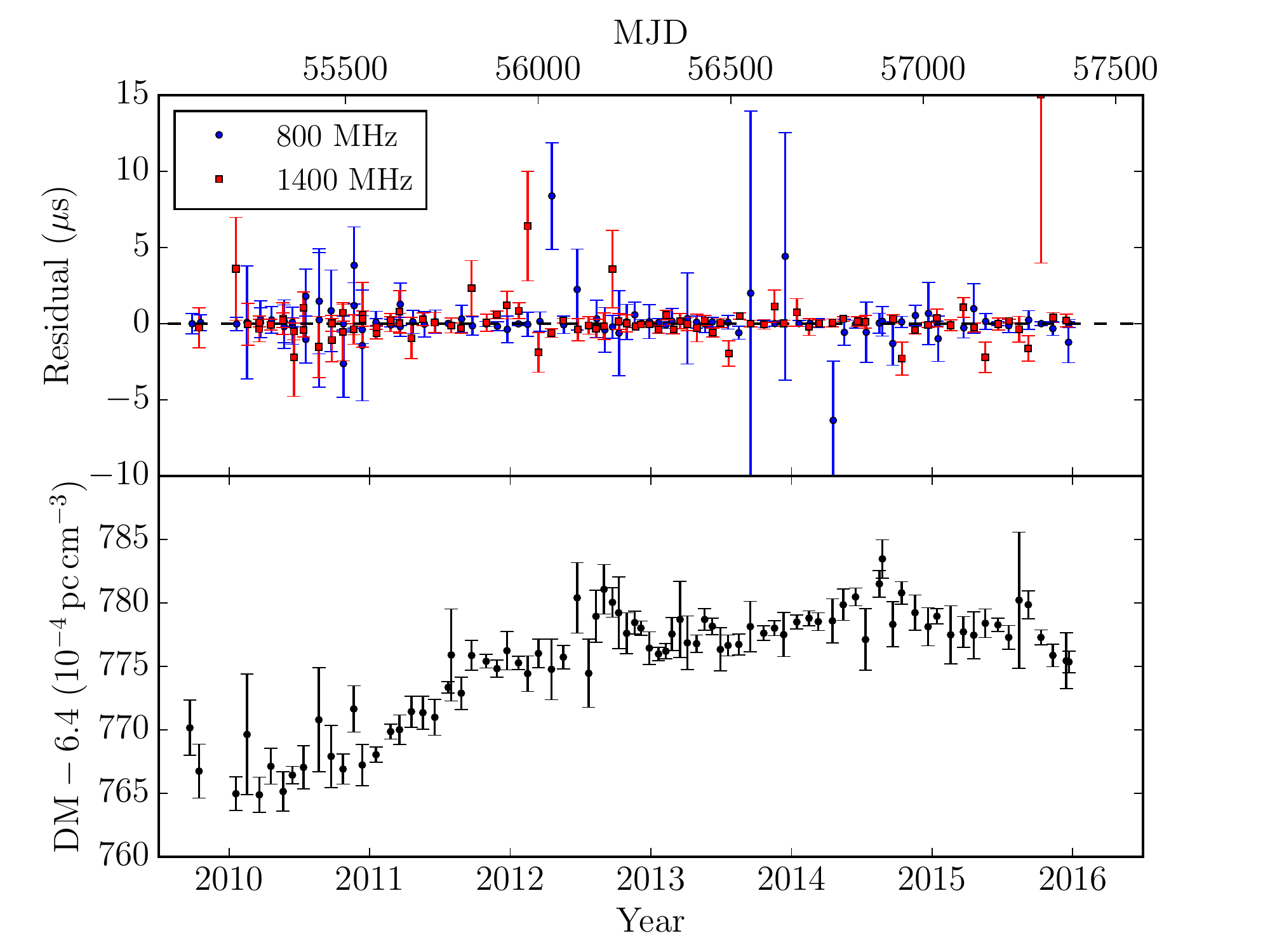}
  \caption{{\it Upper panel.}  Residual pulse arrival times after removing
     the best-fit timing model.  Points represent daily averages for a given
     receiver; blue circles are 800\,MHz and red squares are 1400\,MHz.  
     {\it Lower panel.}  Dispersion measure at every epoch.  These
     values were fit simultaneously with all other timing parameters.  }
  \label{fig:residual_dmx}
\end{figure*}

Best-fit timing model parameters are given in Table~\ref{table:timing}.
The residual pulse arrival times after subtracting off the timing
model, and the variation in DM over time, are shown in Figure~\ref{fig:residual_dmx}.
The two most important results from the timing analysis are
(i) a new measurement of the pulsar parallax
and (ii) a significant measurement of the rotation second frequency
derivative.

The new parallax measurement is $\varpi=0.77\pm 0.23$~mas,
corresponding to a distance of $1.3^{+0.6}_{-0.3}$~kpc.  This agrees
with other recent measurements given in Table~\ref{tab:parallax}.
We checked our distance measurement against the Lutz-Kelker bias-estimate
code of \cite{2012ApJ...755...39V}\footnote{\url{http://psrpop.phys.wvu.edu/LKbias/}}
and found the distance estimate changed by less than $1\sigma$; we elected
not to include this in our reported distance measurement.
In the analysis below, we use our
parallax measurement despite there being values with nominally
smaller uncertainties in the literature, as we believe that 
our dispersion modeling algorithm (fitting independent dispersion
measures at every observing epoch) yields more robust measurements,
especially for pulsars such as PSR~J1024$-$0719 which lie at 
low ecliptic latitudes \citep{2016ApJ...818...92M}.  Adopting other
parallax results would not qualitatively change our analysis.

In the timing model, we parametrized the pulsar spin by the pulsar
rotation frequency $f_0$ and three frequency derivatives ($f_1\equiv df/dt$,
$f_2\equiv d^2f/dt^2$, and $f_3\equiv d^3f/dt^3$). 
The measured
values of these frequency derivatives, and the corresponding
values for pulse period and its derivatives, are listed
in Table~\ref{table:timing}.
Since $f_3$ (or, equivalently, $\dddot{P}$) is of potential interest, we left it in the timing solution even
though its measurement is not formally significant.

We measured a significant rotation frequency second derivative,
$f_2=(-4.1\pm 1.0)\times 10^{-27}~{\rm s}^{-3}$.  This could arise
due to binary motion or due to noise in the rotation of the
pulsar (``timing noise'').  To check for the latter possibility,
we compare our observations with a scaling law for 
timing noise developed by \cite{2010ApJ...725.1607S}.
We use their model which incorporated 
canonical pulsars, millisecond pulsars, and magnetars.
Given the $f_0$ and $f_1$ of \psr, and given the
time span of our observations, 
their model predicts excess residuals of 0.06~$\mu$s,
albeit with large uncertainty.  We estimate 
that our measured $f_2$ would contribute
0.40~$\mu$s if not included in the timing model, substantially 
more than the noise model prediction.
Therefore it is unlikely,
though not impossible, that the observed $f_2$ is due to timing noise.
For the remainder of this paper, we interpret $f_2$ as the jerk, or change 
in acceleration, of the pulsar due to binary motion.  

The measured $f_2$ is equivalent to period second derivative $\ddot{P}=(1.1\pm
0.3)\times 10^{-31}\,{\rm s}^{-1}$.  This is in  agreement with the
value of $(0.70\pm 0.06)\times 10^{-31}\,{\rm s}^{-1}$ reported by
\cite{2016arXiv160105987G}.  Our measurement uncertainty is relatively large
due to covariance between $f_2$ and variations in interstellar DM, which
we fit independently at every epoch simultaneously with the other
parameters; in contrast, \cite{2016arXiv160105987G} used a linear model
in DM which was held fixed in their final timing solutions.  
In the presence of significant DM variations (Figure~\ref{fig:residual_dmx})
we believe our method yields the most robust values of $f_2$
or $\ddot{P}$.  This same reasoning applies to our (non-significant)
measurement of $f_3=(1.1\pm 0.7)\times 10^{-34}\,{\rm s}^{-4}$.  
For example, changing the nature of the DM fit in our 6-year long data-set
from a constant value, to a polynomial of degree up to 7, or to the
by-epoch fit given in Table 2, changes $f_3$ by ${\rm few}\times
10^{-35}\,{\rm s}^{-4}$.  Given that the $f_3$ fit depends on time to the
fourth power, this will be even more apparent in longer data sets.

\begin{deluxetable*}{lc}
\tablewidth{0pt}   
\tablecaption{Timing Parameters of \psr\tablenotemark{a}\label{table:timing}}
\startdata
\cutinhead{Data set}
\rule{0pt}{10pt}%
MJD range         & 55100--57378  \\ 
Data span (yr)         & 6.3 \\ 
Number of TOAs         & 8501 \\
Number of Epochs         & 83  \\
\cutinhead{Timing parameters}
\rule{0pt}{10pt}%
Ecliptic longitude, $\lambda$ (deg.)                            &       160.734351327(14) \\
Ecliptic latitude,  $\beta$ (deg.)                              &       $-$16.04472741(6) \\
Proper motion in $\lambda$, $\mu_\lambda =\dot{\lambda}\,\cos\beta$ (mas\,yr$^{-1}$)   &  $-$14.37(3) \\
Proper motion in $\beta$,  $\mu_\beta =\dot{\beta}$               (mas\,yr$^{-1}$)   &  $-$57.97(13) \\
Parallax, $\varpi$ (mas)                                        &       0.77(23) \\
Rotation frequency, $f_0$ (s$^{-1}$)                            &       193.7156863778085(8) \\
Rotation frequency first derivative, $f_1$ (s$^{-2}$)           &       $-6.9638(4)\times 10^{-16}$ \\
Rotation frequency second derivative, $f_2$ (s$^{-3}$)          &       $-4.1(10)\times 10^{-27}$ \\
Rotation frequency third derivative, $f_3$ (s$^{-4}$)           &       $1.1(7)\times 10^{-34}$ \\
Epoch of period and position (MJD)                              &       56236.000  \\
\cutinhead{Derived quantities}
\rule{0pt}{10pt}%
Right ascension, $\alpha$ (J2000)                                &       $10^{\mathrm h}24^{\mathrm m}38\fs 667384(7)$ \\
Declination, $\delta$    (J2000)                                &       $-07\arcdeg 19\arcmin 19\farcs 5970(2)$ \\
Proper motion in $\alpha$, $\mu_\alpha =\dot{\alpha}\,\cos\delta$ (mas\,yr$^{-1}$)   &  $-$35.26(6) \\
Proper motion in $\delta$, $\mu_\delta =\dot{\delta}$           (mas\,yr$^{-1}$)   &  $-$48.21(13) \\
Total proper motion, $\mu$ (mas\,yr$^{-1}$)       & 59.73(13)  \\
Period, $P$ (s)                                  & 0.00516220456225561(2)  \\
Period first derivative, $\dot{P}$ (s\,s$^{-1}$) & $1.85575(12)\times 10^{-20}$ \\
Period second derivative, $\ddot{P}$  (s$^{-1}$) & $1.1(3)\times 10^{-31}$  \\
Period third derivative, $\dddot{P}$  (s$^{-2}$) & $-3(2)\times 10^{-39}$  \\
Distance, $d_\varpi$ (kpc)                                   & $1.3^{+0.6}_{-0.3}$
\enddata
\tablenotetext{a}{Numbers in parentheses are 1$\sigma$ uncertainties in the last digit quoted.}
\end{deluxetable*}

\section{Optical Observations and Analysis}\label{sec:optical}

\begin{figure*}
  \plotone{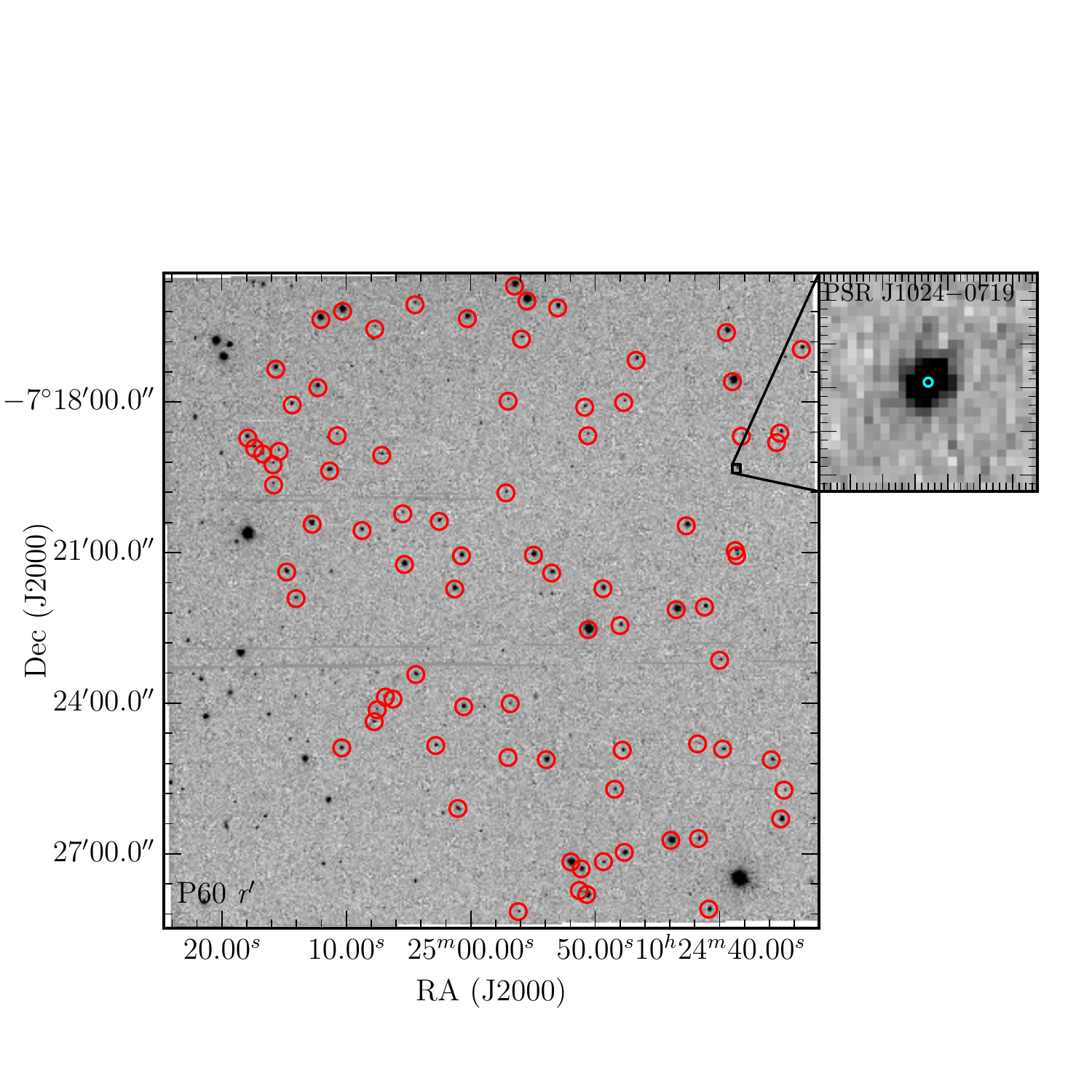}
  \caption{P60 $r^\prime$ image of \psr\ from 2016~Jan~16.  This is a
    single 120\,s exposure.  The 2MASS reference stars used for
    astrometry are circled.  The inset box shows the region around the
    radio position of \psr\ (Table~\ref{table:timing})
    corrected to
    the epoch of the P60 image shown by the cyan circle.  The radius
    of that circle is $0\farcs2$, which is the typical absolute
    astrometric uncertainty of the P60 astrometry.}
  \label{fig:image}
\end{figure*}

\subsection{Optical Imaging}
We obtained images of the field around \psr\ with the Palomar 60-inch
telescope (P60; \citealt{cfm+06}).  The data consist of $4\times 120$\,s
exposures in the $r^\prime$ band on the night of 2016~Jan~16 dithered
by about $20\arcsec$ each.  The data were processed through the
standard P60 pipeline, which determined independent astrometry and
photometric solutions for each image using the USNO B-1.0 catalog. The pipeline is described in full detail in \citet{cfm+06}.

\subsection{Absolute Astrometry}\label{sec:absoluteastrometry}
\label{sec:absolute}
In Figure~\ref{fig:image} we show the position of \psr\ 
(Table~\ref{table:timing})
corrected to the epoch of the P60 images
(MJD~57403).  This position is $0\farcs03$ from the position of
\bright, which we compare with a typical absolute astrometric
uncertainty of $0\farcs2$ for the P60 pipeline.
Likewise, the proper-motion corrected pulsar position is $0\farcs11$
away from Two Micron All-Sky Survey (2MASS; \citealt{2mass}) source
2MASS~10243869$-$0719190.  We
assess the false association rate of the 2MASS source with \psr\ by
considering that within $1\degr$ of \psr, there are 7500 2MASS 
sources brighter than 2MASS~10243869$-$0719190 for an areal density of
$\expnt{(1.84\pm 0.02)}{-4}\,{\rm arcsec}^{-2}$.  Therefore the
association rate is about $\expnt{1.3}{-5}$, and we can be quite
confident that the pulsar is associated with
\bright/2MASS~10243869$-$0719190.

\begin{figure*}
  \plotone{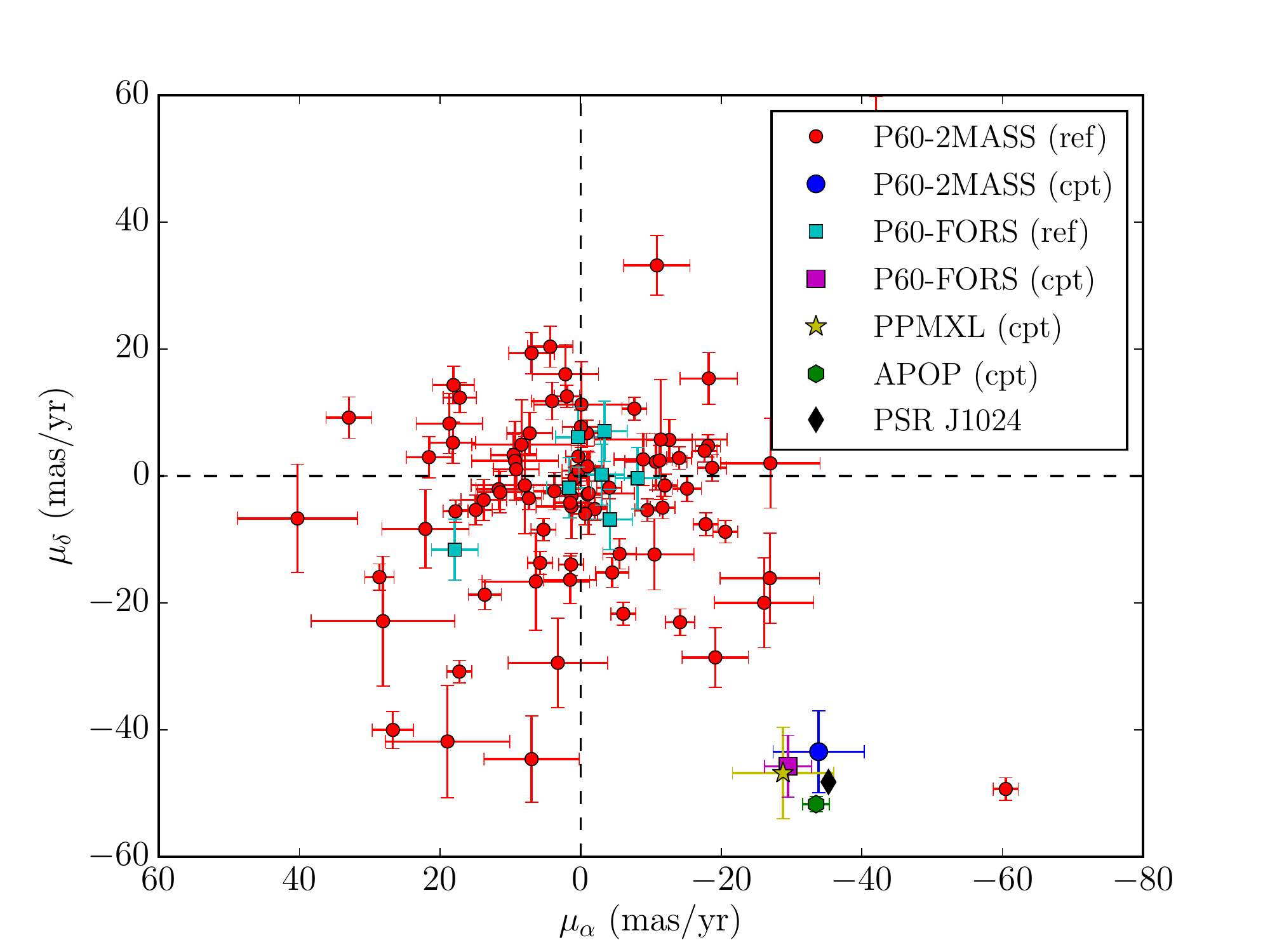}
  \caption{Proper motions for the stars in the \psr\ field.  The
    circles are for the P60 data compared to the 2MASS Point Source Catalog, with the
    red circles the reference sources and the larger blue circle the
    putative counterpart to \psr\ identified by
    \citet{2003A&A...406..245S}.  The squares are for the P60 data
    compared to the archival VLT/FORS1 data, with the cyan squares
    the reference sources and the larger magenta square the putative
    counterpart.  The yellow star is the proper motion of source
    3714292468260686336 from the PPMXL catalog
    \citep{2010AJ....139.2440R}, and the green hexagon the source 39332+0000404 from
    the APOP catalog \citep{2015AJ....150..137Q}, both of which are within $0\farcs15$ of
    \psr\ and consistent with candidate counterpart \bright.  The proper
    motion of \psr\ measured in \S~\ref{sec:timing} is the
    black diamond.}
  \label{fig:pm}
\end{figure*}

We further verify the astrometry by noting that the J2000 position of
the pulsar is 
$0\farcs03$ away from PPMXL
\citep{2010AJ....139.2440R} source 3714292468260686336, and
$0\farcs15$ away from Absolute Proper Motions Outside
the Plane \citep[APOP; ][]{2015AJ....150..137Q} source
APOP~39332+0000404, all of which are consistent with \bright\ (see Table~\ref{tab:astrom}).
For 2MASS and PPMXL the proposed counterpart is within 1-$\sigma$ of
the proper motion-corrected radio timing position.  In APOP the
proposed counterpart is slightly further away and the quoted accuracy
of APOP of $\pm 0.2\,$mas relative to the ICRF suggests that the
offset, $0\farcs 15\pm 0\farcs 03$,  may be significant, but we are cautious with frame
ties between the radio and optical systems (e.g.,
\citealt{2016arXiv160208868V}) so in what follows we largely treat
this as an upper limit to the projected separation.

\begin{deluxetable*}{l c c c c l}
 \tablewidth{0pt}
\tablecaption{Astrometry of \psr\ and its Optical Counterpart\label{tab:astrom}}
\tablehead{
 \colhead{Survey} & \colhead{$\Delta \alpha$\tablenotemark{a}} & \colhead{$\Delta \delta$\tablenotemark{a}} &
 \colhead{$\mu_\alpha$} & \colhead{$\mu_\delta$} & Reference\\
 & \colhead{(arcsec)} & \colhead{(arcsec)} & \colhead{(mas\,yr$^{-1}$)}& \colhead{(mas\,yr$^{-1}$)}}
\startdata
NANOGrav & \multicolumn{1}{c}{0} & \multicolumn{1}{c}{0} & $-35.26\pm0.06$ &
$-48.21\pm0.13$ & This paper \\
P60-2MASS & \phs$0.03\pm0.20$ & $-0.11\pm0.20$ &  $-33.9\pm6.5$ &
$-43.4\pm6.5$ & \citet{2mass}/this paper\\
APOP & $-0.11\pm0.04$& \phs$0.09\pm0.03$ & $-33.5\pm1.9$ & $-51.7\pm1.2$  & \citet{2015AJ....150..137Q}\\
PPMXL & \phs$0.00\pm0.11$ & \phs$0.03\pm0.11$ & $-29.0\pm7.0$ & $-44.2\pm7.0$ & \citet{2010AJ....139.2440R}\tablenotemark{b}\\
\enddata
\tablenotetext{a}{Offsets between the optical counterpart and the
  radio position of \psr\ at epoch 2000.0}
\tablenotetext{b}{PPMXL proper motions have been corrected according
  to \citet{2016arXiv160208868V}.}
\end{deluxetable*}

\subsection{Relative Astrometry and Proper Motions}
\label{sec:propermotion}
To determine proper motions of the stars in this field, we compared
our P60 observations against the 2MASS Point Source Catalog (PSC). We
measured the positions of all of the stars in the P60 images using
\texttt{sextractor} \citep{ba96}, and matched each exposure separately
to the 2MASS sources.  We required that the source be $<3\arcsec$ from
its 2MASS position and that it have no quality flags suggesting
questionable data (bad pixels, saturation, etc), and found 82
reference sources over the P60 image plus the possible counterpart to
the pulsar (which is itself a 2MASS source as discussed above); see
Figure~\ref{fig:image}.  We then computed position offsets between the
positions measured in the P60 images (MJD~57403) and the 2MASS PSC
(MJD~51193), which we show in Figure~\ref{fig:pm}.

The majority of the reference stars had proper motions with amplitudes
$<20\,\masyr$ and were clustered around 0. A few stars had
individually significant proper motions, among them
\bright.  We find a proper motion for
this star of $(-34\pm6\,\masyr, -43\pm6\,\masyr)$ which is within
1-$\sigma$ of our measurement of the pulsar proper motion
(Table~\ref{table:timing}).

We verified this proper motion using the same imaging data used by
\citet{2003A&A...406..245S}.  We retrieved data taken by the ESO 8.2m
Very Large Telescope Antu (VLT-UT1) with the FORS1 CCD in the
narrow-field imaging mode in the Bessel $V$ band (the other bands were
similar) from the ESO archive, finding $3\times 120\,$s exposures.  We
reduced the data using custom routines, removing the overscan,
subtracting a bias frame, and then flat-fielding the images.  The much
narrower field-of-view ($205\arcsec$ vs.\ $774\arcsec$) and the much
larger telescope means that many fewer reference stars were available,
with only 7 sources that we could match to our P60 data.  We
determined the  position offset of all of the sources in the FORS1
data (MJD~51996) compared to the  P60 data, averaging over the
individual exposures in both data-sets.  We find a proper motion of
2MASS~J10243869$-$0719190 to be $(-29\pm4\,\masyr,-45\pm4\,\masyr)$
which is consistent with both our measurement from P60 to 2MASS as
well as the NANOGrav proper motion (Fig.~\ref{fig:pm}).  There were insufficient sources
that matched between the FORS1 data and 2MASS for a third proper
motion measurement, as well as a significantly reduced time baseline
(800\,d, vs.\ 15--17\,yr).

Finally, the proper motions of the radio pulsar and
the APOP and PPMXL sources (Table~\ref{tab:astrom} and
Fig.~\ref{fig:pm}) are all consistent to within the uncertainties.  We
conclude that \psr\ and \bright\ form a common proper motion pair.

\subsection{Spectral analysis}
\label{sec:spec}
Optical spectra of \bright\ were obtained with the Palomar 200-inch
telescope and the Double-Beam Spectrograph (DBSP; \citealt{og+82}) on
2016 Jan 30 using a low resolution mode ($R\sim1500$).  We took three
exposures with an exposure time of $1000$\,s each.  Both arms of the
spectrograph were reduced using a custom \texttt{PyRAF}-based
pipeline.\footnote{https://github.com/ebellm/pyraf-dbsp} The pipeline
performs standard image processing and spectral reduction procedures,
including bias subtraction, flat-field correction, wavelength
calibration, optimal spectral extraction \citep{1986PASP...98..609H},
and flux calibration.  For the analysis all three individual exposures
were combined resulting in a SNR of about 25 at 7000\,\AA.

\begin{figure*}
  \plotone{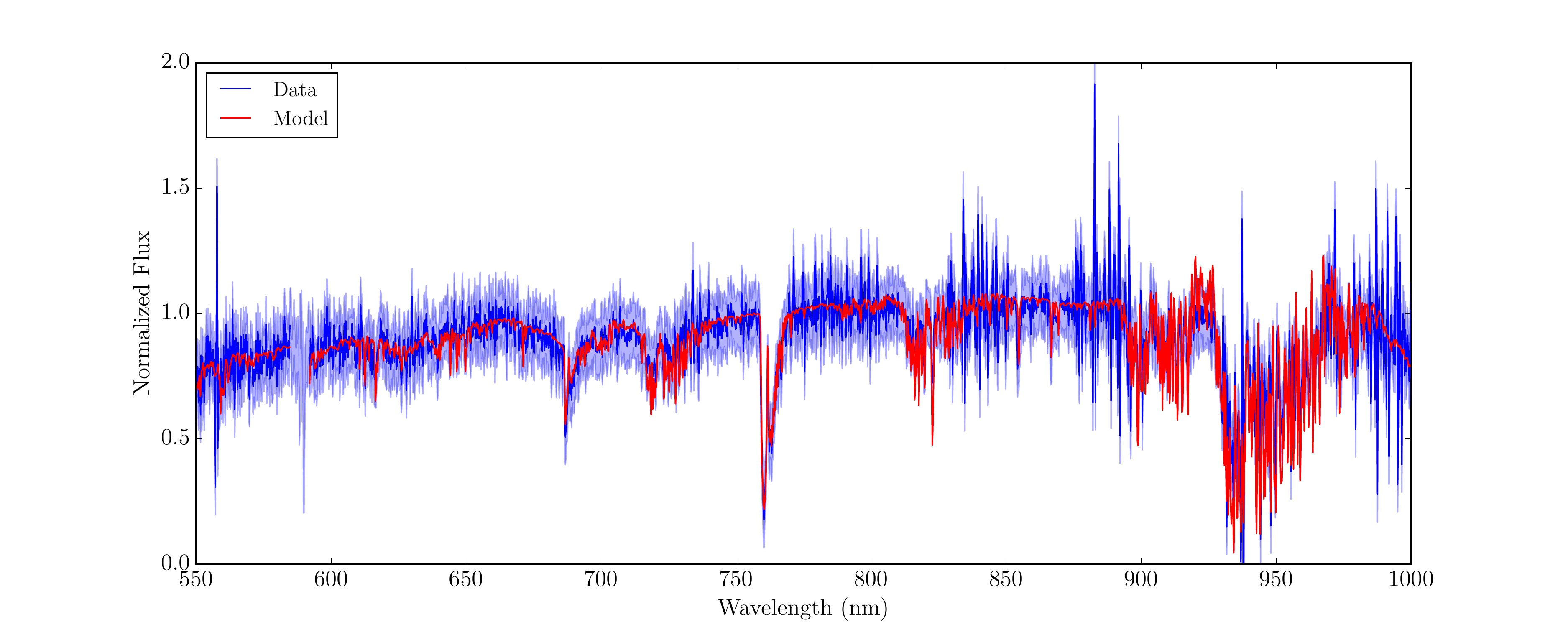}
  \caption{Normalized spectrum of \bright\ from the Palomar 200-inch
    DBSP observation.  The blue line is the data, with the
    uncertainties indicated by the shaded region.  The red line is the
    best-fit Phoenix model with radial velocity $v_r = 221$\,\kms,
    effective temperature of $T_{\mathrm{eff}} = 3900$\,K, and a
    metallicity of $Z = -0.84$\,dex (the surface gravity was
    fixed to 4.9\,dex).}
  \label{fig:spectrum}
\end{figure*}

We fit the red part of the normalized spectrum using Phoenix models
\citep{2013A&A...553A...6H}, which are multiplied with a telluric
transmission spectrum \citep{2014A&A...568A...9M} to account for
telluric absorption.  The region around the \ion{Na}{1} doublet
(5889.961\,\AA--5895.924\,\AA) was ignored because of contamination
with night sky emission lines. The telluric absorption bands were used
to correct the wavelength scale for instrument flexure. The fitting
parameters included the radial velocity $v_r$, the effective
temperature $T_{\mathrm{eff}}$, and the metallicity $Z$.  Since
spectroscopic determination of surface gravities for cool stars is
notoriously difficult even from high-resolution, high-fidelity spectra
\citep{2014AA...570A.122S}, we kept  the
surface gravity  fixed at $\log g = 4.9$\,dex (see
\S~\ref{sec:sed}). We found a good fit (Fig.~\ref{fig:spectrum}) with
a heliocentric velocity of $v_r = 221\pm30$\,\kms, an effective
temperature of $T_{\mathrm{eff}} = 3900^{+60}_{-40}$\,K (spectral type
of roughly M0), and a metallicity of $Z = -0.84^{+0.10}_{-0.09}$\,dex
(uncertainties are single-parameter $1\sigma$-confidence intervals
based on the $\chi^2$ statistics after it was re-scaled to yield a
reduced $\chi^2$ of about $1$, and the radial velocity uncertainty
includes systematic uncertainties to account for the wavelength
scale).

\subsection{Spectral Energy Distribution}
\label{sec:sed}
\begin{figure*}
  \plotone{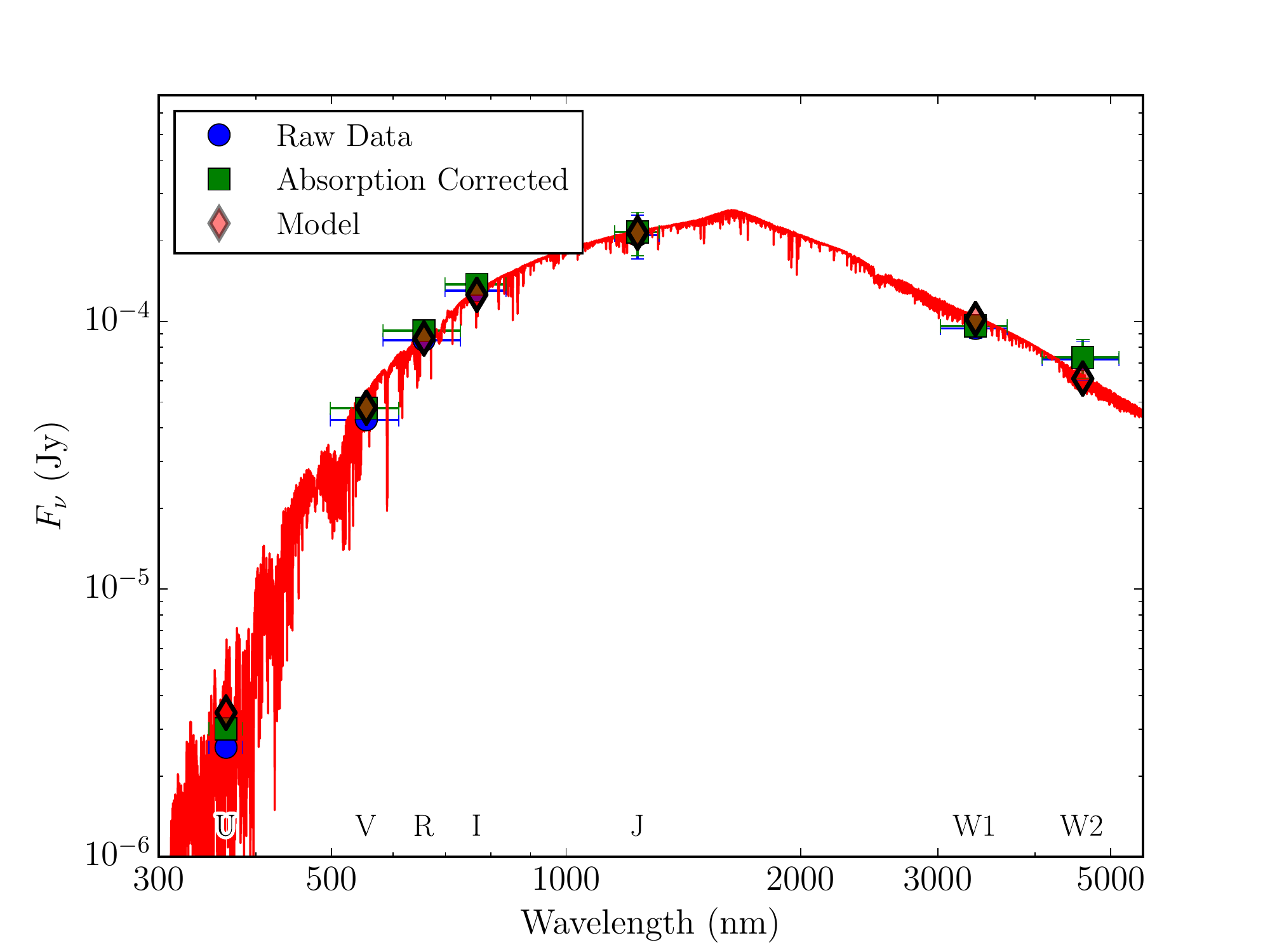}
  \caption{Spectral-energy distribution (SED) of the \bright.  The
    blue circles are the raw photometry from
    \citet{2003A&A...406..245S} along with archival 2MASS and
    \textit{WISE} data, where we use the zero-point flux densities
    from \citet{bcp98}, \citet{2003AJ....126.1090C}, and
    \citet{2011ApJ...735..112J}.  The green squares have been
    corrected for extinction with $E(B-V)=0.04$.  The red curve is a
    Phoenix model atmosphere for the best-fit effective temperature of
    3850\,K, and the red diamonds are that model atmosphere integrated
    over the filter passbands.}
  \label{fig:sed}
\end{figure*}

Based on the spectroscopic result, we analyzed the spectral energy
distribution using the archival FORS1 photometry from
\citet{2003A&A...406..245S} along with the 2MASS $J$-band (the source
was not detected in the $K_s$-band, and the $H$-band point had low
signal-to-noise) and \textit{WISE} \citep{2010AJ....140.1868W} $W_1$-
and $W_2$-band data again using Phoenix models.  For all observed
magnitudes, a systematic uncertainty of $0.045$\,mag is added in
quadrature to have a reduced $\chi^2$ of about $1$ at the best fit.
We determined the line-of-sight reddening using the three-dimensional
models of \citet{2015ApJ...810...25G}, finding $E(B-V)=0.04$\,mag for
distances $>1$\,kpc (consistent with the value used by
\citealt{2003A&A...406..245S}).  With the metallicity and surface
gravity set to the spectroscopic result, we obtain fit parameters very
similar to the spectroscopic values: $T_{\mathrm{eff}} =
3874^{+208}_{-\phantom{0}29}$\,K and a distance (based on an assumed surface gravity of $\log g = 4.9$\,dex and mass $M = 0.4\,M_\odot$, appropriate for a low-metallicity star with
this effective temperature; \citealt{1997A&A...327.1039C}) of
$d=1.08\pm0.04$\,kpc, consistent with the radio timing.  The fit is
shown in Figure~\ref{fig:sed}.

\section{A Wide Binary Companion?}
\label{sec:binary}
Following the discussions in \citet{2016ApJ...818...92M} and
\citet{2016arXiv160208511D}, we consider whether or not \psr\ and
\bright\ form a binary and, if so, how we could constrain its
parameters (also see \citealt{2015MNRAS.451..581L}).  We have shown
that the pulsar and the optical source have absolute positions
consistent within uncertainties (\S~\ref{sec:absolute}). If we adopt
the recent parallax distances for \psr, rather than the DM distance,
then its distance is also consistent with the main-sequence distance
for \bright.  Therefore the objects appear to align in three
dimensions.  Since they also form a common proper motion pair, they
align in two further dimensions of phase space.  Could a wide binary
system satisfy our further dynamical constraints?  We consider three
specific constraints.
\begin{enumerate}
  \item The intrinsic period derivative of the pulsar should be $>0$,
    and is likely $\lesssim 10^{-19}\,{\rm s\,s}^{-1}$ consistent with
    most MSPs.
  \item The pulsar and putative companion are separated by
    $\lesssim 0\farcs15$ on the sky.
  \item The pulsar should have a period second derivative $\ddot
    P=\expnt{(3.4\pm0.9)}{-24}\,{\rm yr}^{-1}$ (\S~\ref{sec:timing}).
\end{enumerate}

\subsection{Circular Orbit Models}\label{sec:circularorbits}

While there is no {\it a priori} reason to assume the orbits of the
pulsar and companion are circular, such an assumption simplifies
the analysis and can elucidate the broad properties of the system.
Thus we begin by considering circular orbits; we broaden the
analysis to include eccentric orbits in \S\ref{sec:eccentricorbits}.

{\it Constraint \#1.}
If we posit that the pulsar and companion are in a wide orbit such
that only low-order period derivatives are apparent in the timing
residuals, we can constrain the properties of the orbit.   First,
we take:
\begin{equation}
\dot P_{\rm int} = \dot P_{\rm obs} - \dot P_{\rm Shk} -
\dot P_{\rm orb}
\label{eqn:pdotint}
\end{equation}
as the intrinsic $\dot P_{\rm int}$, where we correct the observed $\dot P_{\rm obs}$ for
the Shklovskii effect, $\dot P_{\rm Shk}$, and for any dynamical 
influence of an orbit, $\dot P_{\rm orb}$.
Note that $\dot P_{\rm orb}$ refers to a
change in the pulsar period due to orbital motion, not a change in the
period of the orbit.
We ignore corrections for differential acceleration in the Galactic
potential, which are small for \psr. 

\begin{figure*}
  \plotone{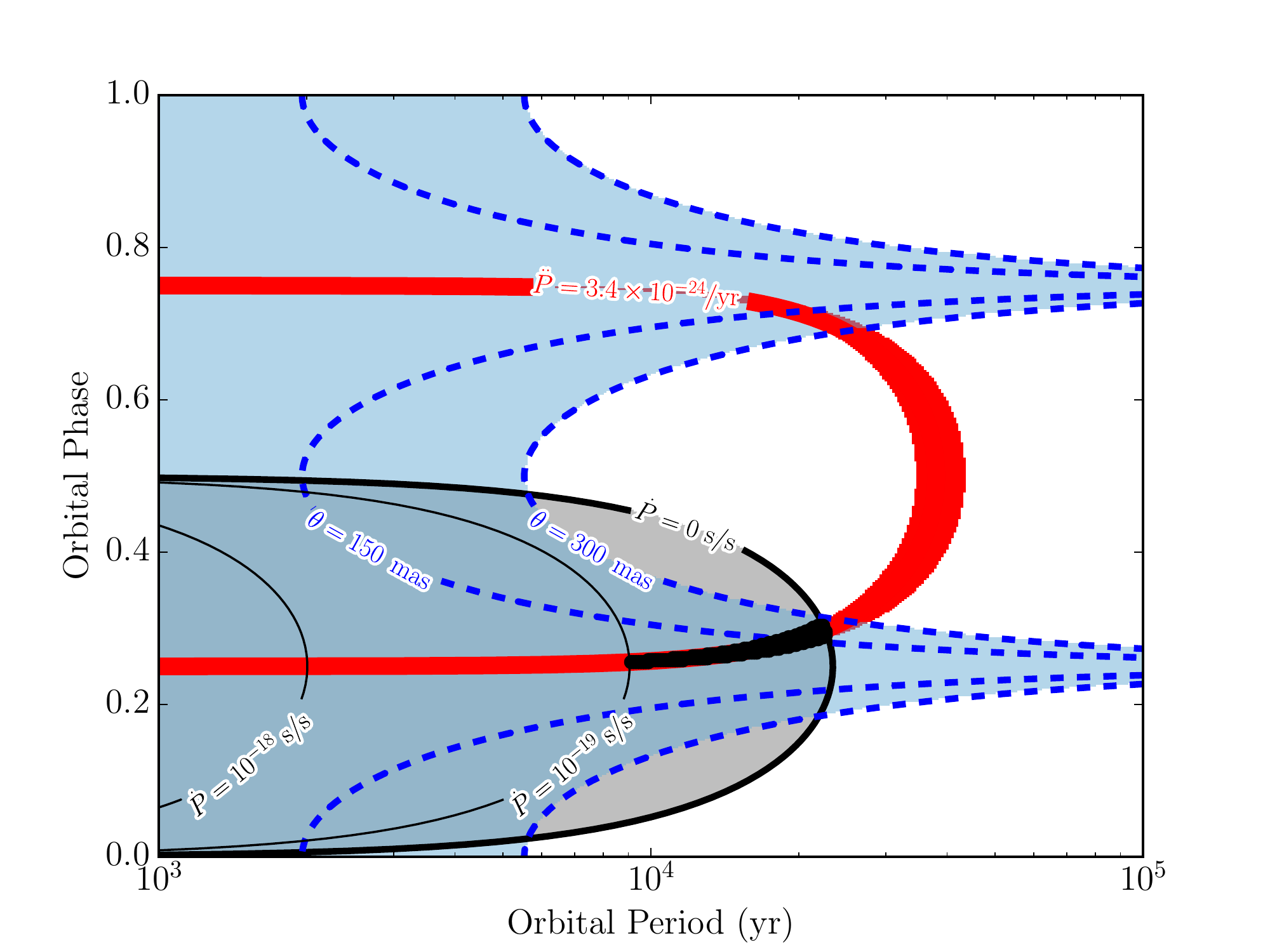}
  \caption{%
    Constraints on the \psr\ orbital period and orbital phase
    (mean anomaly), assuming a circular, edge-on orbit.
    The gray shaded region shows the constraint $\dot{P}_{\rm int}<0$, and
    solid black lines show constraints at
    specific values of $\dot{P}_{\rm int}$.  Typical 
    millisecond pulsars have $0<\dot{P}_{\rm int}<10^{-19}\,{\rm s}\,{\rm s}^{-1}$,
    so the \psr\ orbit should lie between these lines.
    The blue shaded region shows the constraint from angular separation
    on the sky, $\theta<300$~mas,
    and dashed blue lines show constraints at specific values of $\theta$.
    The red contour is where $\ddot P=\expnt{(3.4\pm0.9)}{-24}\,{\rm yr}^{-1}$.
    The thick black region meets all of these criteria.
    }    
  \label{fig:pdot}
\end{figure*}

For
a circular orbit,
\begin{eqnarray}
  \dot P_{\rm orb}& =& \frac{G M_{\rm c} P \sin i}{a^2c}\sin \phi \nonumber \\
  &=& \frac{G^{1/3}
    M_{\rm c}P \sin i}{(M_{\rm c}+M_{\rm psr})^{2/3}c}
  \left(\frac{2\pi}{P_{\rm b}}\right)^{4/3} \sin\phi,
\label{eqn:pdotorb}
\end{eqnarray}
where $M_{\rm c}$ is the companion mass, $M_{\rm psr}$
is the pulsar mass,
$a$ is the orbital semi-major axis
(full separation between the pulsar and companion), $i$ is the
inclination, $P_{\rm b}$ is the binary period, and $\phi$ is the orbital
phase (mean anomaly, measured from 0 to 1, with 0 being the ascending
node).  
Figure~\ref{fig:pdot} (black lines) shows the constraints
that arise from equations~\ref{eqn:pdotint} and~\ref{eqn:pdotorb}
for different values of $\dot{P}_{\rm int}$, using our observed
value of the companion mass, $M_{\rm c}\approx 0.4\,M_\odot$,
and assuming (for simplicity)
an edge-on orbit, $i=90^\circ$, and a fixed pulsar mass of
$M_{\rm psr} = 1.54\,M_\odot$
\citep{2016arXiv160302698O}. 
Typical solutions have orbital
periods of $P_{\rm b}\approx 10$\,kyr, with the maximum allowed value
of $\approx 30$~kyr.  In order to have a positive $\dot{P}_{\rm int}$,
the pulsar must have orbital phase $0.0<\phi<0.5$.

{\it Constraint \#2.}  We consider the projected separation between
\psr\ and the putative companion.  For a wide orbit, there will be
some phases where the projected separation between the pulsar and the
companion is quite large.  The constraints for the parameters of the
\psr\ system are shown in Figure~\ref{fig:pdot} as dashed blue lines
for our estimated upper limit on separation, $\theta<150$~mas
(\S\ref{sec:absoluteastrometry}) and for a more conservative
$\theta<300$~mas.  For a circular orbit, the maximum separation is
$919 (P_{\rm b}/20\,{\rm kyr})^{3/2} d_1\,$mas at quadrature ($\phi=0$
or $\phi=0.5$), where the distance is $1\,d_1\,$kpc, while the minimum
projected separation is $919\cos i (P_{\rm b}/20\,{\rm kyr})^{3/2}
d_1\,$mas at conjunction ($\phi=0.25$ or $\phi=0.75$).
So if the pulsar and companion were near quadrature they would violate
our limit on $\theta$ regardless of inclination, but near conjunction
they can satisfy this constraint.

{\it Constraint \#3.}
Finally we consider the period second derivative.  This comes from the
jerk (time derivative of the acceleration) along the line-of-sight in
the orbit.  In a circular orbit, the dynamical $\ddot P$ is
\begin{eqnarray}
  \ddot P_{\rm orb}& =& \frac{G^{3/2} M_{\rm c} (M_{\rm c}+M_{\rm psr}) P \sin i}{a^{7/2}
    c}\cos\phi \nonumber \\
  &=&\frac{ G^{1/3} M_{\rm c} P \sin i}{(M_{\rm c}+M_{\rm psr})^{2/3}
    c}\left(\frac{2\pi}{P_{\rm b}}\right)^{7/3}\cos\phi.
\end{eqnarray}
The constraint based on the observed $\ddot P$ is shown in Figure~\ref{fig:pdot}.

As seen in Figure~\ref{fig:pdot}, all three of these
constraints are satisfied by edge-on circular binary
systems with orbital periods
10--30\,kyr and appropriate orbital phases.
For inclined circular orbits (not shown) these
constraints can still all be met.  The orbital period decreases to around $2\,$kyr as
the inclination approaches 0.

\subsection{Eccentric Orbit Models}\label{sec:eccentricorbits}

We can find solutions for the general case of inclined, eccentric
orbits (based on \citealt{1997ApJ...479..948J,2001MNRAS.322..885F}).
To fully explore the phase space, we undertook a Markov-Chain Monte Carlo
(MCMC) exploration of the 8-dimensional phase space.  We varied orbital period $P_{\rm b}$,
inclination $i$, eccentricity $e$, distance $d$, companion mass $M_{\rm c}$,
and proper motion $\mu$, along with nuisance parameters for the mean
anomaly and the longitude of periastron.  As in \S\ref{sec:circularorbits}, 
for simplicity, we held the pulsar mass 
fixed at $1.54\,M_\odot$; analysis with different pulsar mass values would
produce qualitatively similar results, with minor rescalings of parameter
values.  We assumed prior distributions on the
parallax ($\varpi=0.78\pm0.23\,$mas) and
proper motion ($\mu=59.73\pm0.13\,\masyr$) from our updated timing
(\S~\ref{sec:timing}), and
$M_{\rm c}=0.4\pm0.1\,M_\odot$ to match our SED fitting.  We also included
flat prior distributions on $\cos i$ and $\log P_{\rm b}$.  The posterior
was evaluated with a hard cutoff for $\dot P_{\rm int}$,
requiring it to be between 0 and $10^{-19}\,{\rm s\,s}^{-1}$.  We
evaluated goodness-of-fit by comparing the inferred $\ddot P$ against
the measured value of $\expnt{(3.4\pm0.9)}{-24}\,{\rm yr}^{-1}$ as well
as the projected separation with best-fit value of 0 and uncertainty
of $0\farcs15$.
Using \texttt{emcee} \citep{fmhlg13} we used 600 ``walkers'' for
50,000 iterations each, starting the walkers off randomly distributed
in phase space according to the priors described above.  After
removing 100 iterations for ``burn-in'' and thinning the samples by a
factor of about 1000 to account for correlations among the points, we
had roughly 20,000 individual samples for each parameter.  The results
are shown in Figure~\ref{fig:mcmc}.  We see results broadly consistent
with our inferences from the edge-on circular orbits: binary periods
near $10^4$\,yr are preferred, as are edge-on orbits, and overall
lower eccentricities are better but no eccentricity is excluded.
There is a general covariance between $P_{\rm b}$ and $i$, with smaller
periods needed at more face-on inclinations (reinforcing our results
from above) but allowing larger distances, and the minimum binary
periods are around 2\,kyr.  The lower binary periods are preferred
solutions with higher eccentricities, and there is a clear selection
of eccentricity based on the sign of $\dddot P$: if $\dddot P>0$ then
more circular (and hence wider and more edge-on) orbits are preferred,
but if $\dddot P<0$ then circular orbits cannot fit the data
(following Fig.~\ref{fig:pdot}) and we need higher eccentricities,
lower $P_{\rm b}$, and more face-on orbits.

Overall, we conclude that a wide binary system is completely compatible
with all of the observational constraints on \psr\ and 1024-Br.

\section{Discussion}
\label{sec:discuss}
We now consider the implications of such a binary system for some of
the puzzling measurements discussed above.  

The $\gamma$-ray efficiency should be revised for the updated $\dot P$ and
distance.  The $\gamma$-ray flux is $\expnt{3.8}{-12}\,{\rm
  erg\,s}^{-1}\,{\rm cm}^{-2}$ \citep{2013MNRAS.430..571E}, so the
luminosity is $\expnt{4.5}{32}d_1^2\,{\rm erg\,s}^{-1}$ (assuming
beaming into $4\pi$\,ster).   
If $\dot P_{\rm int}$ is as high as $10^{-19}\,{\rm s\,s}^{-1}$
which is certainly possible (Figs.~\ref{fig:pdot} and \ref{fig:mcmc}),
this implies a spin-down luminosity as large as $\dot
E=\expnt{3.1}{34}\,{\rm erg\,s}^{-1}$, or a $\gamma$-ray efficiency as
low as $1.5 d_1^{2}$\%.  Likely the true value is not this low, but
this at least resolves the possible problem raised by
\citet{2016ApJ...818...92M} regarding the apparent extremely
high efficiency at the parallax distance.

Similarly, we must revise the analysis of the X-ray luminosity.
\citet{2006ApJ...638..951Z} find a thermal luminosity of
$\expnt{2.6}{30}d_1^2\,{\rm erg\,s}^{-1}$.  If $\dot E$ is as high as
that in the previous analysis, the X-ray efficiency would be as low as
$10^{-4.1}$, which is somewhat lower than most objects in
\citet{2006ApJ...638..951Z} but less discrepant than it was before.

\begin{figure*}
  \plotone{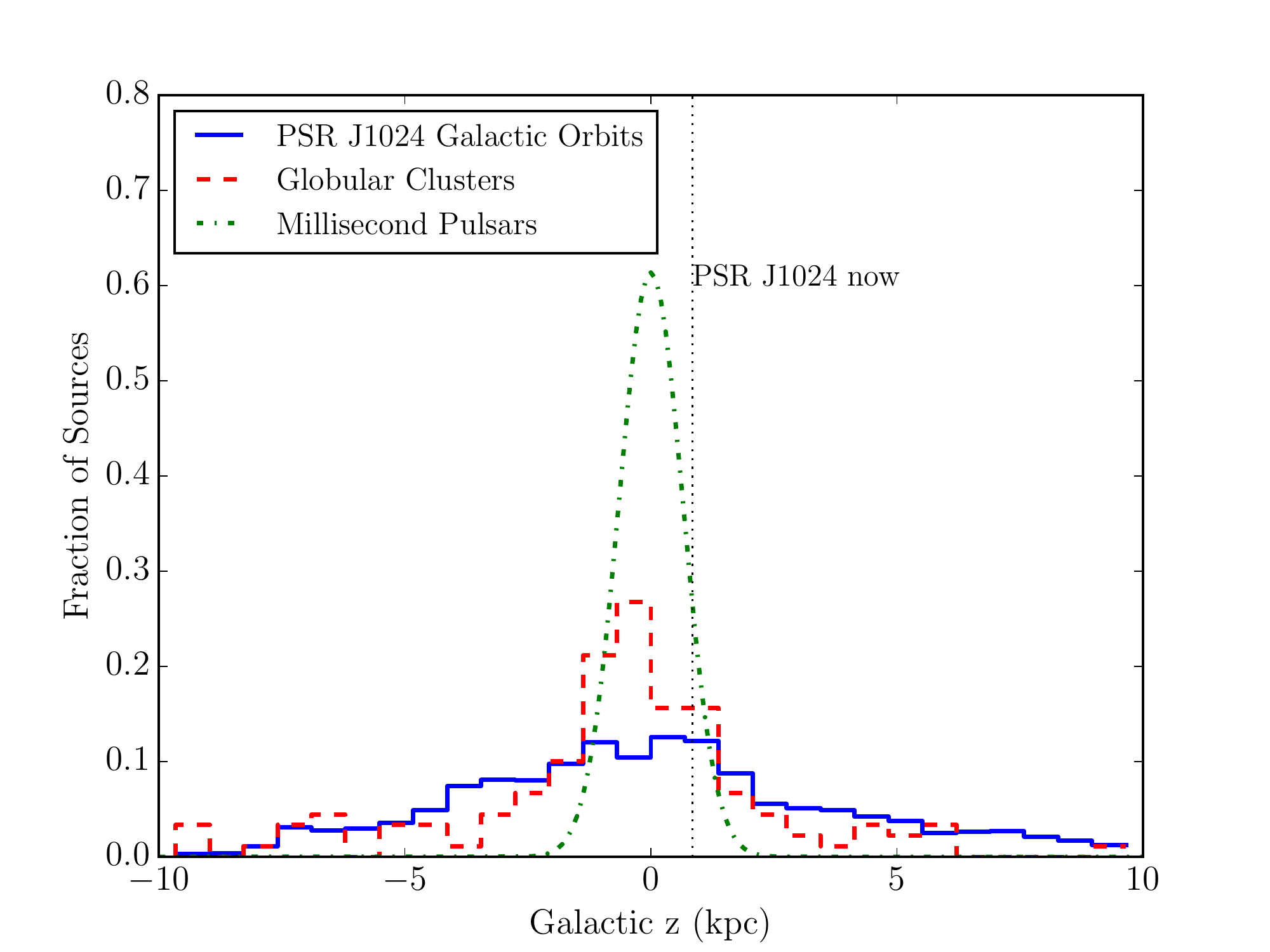}
  \caption{Distribution of distance of \psr\ above/below the Galactic
    plane $z$ (solid blue line), shown for 100 orbits in the Galactic
    potential over the past 1\,Gyr calculated using \texttt{galpy}
    \citep{2015ApJS..216...29B}.  We also compare with the
    distribution of Galactic globular clusters at their current
    positions \citep[][2010 edition; red dashed
      line]{1996AJ....112.1487H} and the vertical distribution of MSPs
    from \citet[][green dot-dashed line]{cc97}.  The current $z$ of
    \psr\ (0.84\,kpc) is the vertical dotted line.  }
  \label{fig:galactic}
\end{figure*}

While a wide binary system resolves some of the puzzles regarding the
distance, a major remaining puzzle is its high transverse velocity,
$v_{\perp}=282 d_1\,\kms$, and what that implies about the possible
formation mechanisms.  As discussed by \citet{2016ApJ...818...92M}, if
placed at its parallax distance, \psr\ has a much higher velocity than
most MSPs. Using a radial velocity $v_r=221\pm30\,\kms$, we find
 velocities
$(U,V,W)=(-82\pm15,-436\pm122,-164\pm135)\,\kms$ relative to the Local
 Standard of Rest using the Solar
motion from \citet{hbrj05}.  This agrees roughly with the velocity
ellipsoid for metal-poor halo stars \citep[e.g.,][although it prefers
  metallicities $\lesssim -2$\,dex]{2000AJ....119.2843C}, or with the
radial \citep[][2010 edition]{1996AJ....112.1487H} and tangential
\citep[e.g.,][]{2007ApJ...657L..93K} velocities of globular clusters.
However, it is about 4 times the velocity dispersion for MSPs
\citep{cc97,2016ApJ...818...92M}, and if we integrate the orbit of
\psr\ in the Galactic potential (using \texttt{galpy};
\citealt{2015ApJS..216...29B}) we find a scale height of 2--4\,kpc
(Fig.~\ref{fig:galactic}), compared to $0.65\,$kpc for MSPs
\citep[][also see \citealt{2013MNRAS.434.1387L}]{cc97}.  This suggests
that kinematically, \psr\ belongs to a separate population than the
vast majority of MSPs, and this may relate to how it was formed.

Young pulsars with very high space velocities
are known \citep[e.g.,][]{2005ApJ...630L..61C}, and they likely rely
on binary disruption and/or supernova kicks for their high velocities.
Similarly, hypervelocity stars \citep[e.g.,][]{2005ApJ...622L..33B}
are often thought to originate \citep{2015MNRAS.448L...6T} from
binaries disrupted by a supermassive black hole
\citep[e.g.,][]{1988Natur.331..687H} or a supernova
\citep{1961BAN....15..265B}; other possibilities such as a tidal
stream \citep{nemeth16} or dynamical ejection following an exchange in
a dense stellar environment \citep{1974A&A....35..237A} may also
operate.  However, the case of \psr\ is different from both young
pulsars and hypervelocity stars, in that it is presumably recycled
following prolonged stellar evolution in a close ($P_{\rm b}\sim$day) binary
\citep[e.g.,][]{tlk12} with a companion that is now absent.  Instead
its companion is anomalous, more like the eccentric binary
\object[PSR J1903+0327]{PSR~J1903+0327} \citep{crl+08,2011MNRAS.412.2763F}.  We
note that estimates suggest less than 1\% of the MSP population
originates from the halo \citep{cc97}, which could be consistent with
finding a single object like \psr\ in the $\sim$hundred well-timed
MSPs, but \psr\ likely requires a denser natal
environment such as a globular cluster \citep[cf.][]{nemeth16} to have
had the dynamical encounters that removed the original companion and
left the current one.

To further explore this topic we compare with \object[PSR
  B1620-26]{PSR~B1620$-$26} \citep{1988Natur.332...45L} in the
globular cluster M4, which has a white dwarf in a 191-day orbit and a
Jupiter-mass companion in a decades-long outer orbit \citep{tacl99}.
Most formation scenarios favor a dynamical encounter in the dense core
of the globular cluster \citep{2003Sci...301..193S} which exchanged
the planet into the MSP system to explain the wide eccentric orbit.
Recoil following the exchange can explain why the PSR~B1620$-$26
system is currently on the outskirts of M4 on a wide orbit in the
cluster's potential, although it is still likely bound.  As much as
50\% of the globular cluster MSP population could be ejected
\citep{2008MNRAS.386..553I}, and further objects could be tidally
stripped \citep{1997ApJ...474..223G}, which could explain the origin
of \psr\ in the Galactic plane \citep[cf.][]{crl+08}.

We suggest that \psr\ was formed in a globular cluster (which form
MSPs at a very high rate due to the many stellar encounters; e.g.,
\citealt{1995ApJS...99..609S,2014A&A...561A..11V}), and that its
initial evolution was much like most other such systems with recycling
in a compact binary with a white dwarf.  A subsequent dynamical
encounter with another binary \citep[also see][]{2015ApJ...807L..23D}
exchanged/ejected the white dwarf and led to the current system.
There might have also been a phase including a triple system, whose
disruption might explain the very wide orbit.
Eventually, either
as the result of the initial encounter or subsequent encounters the
\psr\ system would have been ejected from the globular cluster (which
only requires a recoil velocity of $\sim 30\,\kms$, consistent with
most dynamical predictions). The velocity of the system now would be
the halo velocity of the cluster plus a small amount, consistent with
the orbit we now see.  Note that we cannot trace back the system to a
potential cluster of origin given the poor knowledge of space
velocities for most globular clusters and the unknown age of this
system.
However, the sub-solar metallicity we see for \bright\ is consistent
with typical values for globular clusters.

\cite{2016ApJ...818...92M} analyzed the MSP velocity
distribution and posited a model in which the bulk of the
MSP population is formed in the Galactic disk and has velocities 
similar to the thermal velocities of other old stellar populations, 
but in which there are a few high-speed outliers.  Our formation scenario
for \psr\ suggests that ejecta of globular clusters may be the source
of the outlier population.

\citet{bjs+16} came to conclusions very similar to ours regarding the nature of the
\psr\ system using largely independent radio and optical data-sets.  They
proposed a formation scenario in which the system is the remnant of a
hierarchical triple system formed in the Galactic disk, with its high space
velocity the result of a supernova kick.  In both scenarios some degree of
fine tuning is required to end up with the current barely-bound binary and to
match the space velocity.
The true origin may be a combination of both
scenarios, with a hierarchical triple evolving in a globular cluster and
being ejected as it evolves into a wide MSP binary system.
Such a scenario might remove some of the fine tuning needed above and in
 \citet{bjs+16}.

\section{Conclusions}\label{sec:conclusions}
We have presented new radio timing along with archival optical data
that strongly suggest \psr\ is in a wide (2--20\,kyr) binary orbit
with a low-mass stellar companion.  Our preferred formation mechanism
is that the system was formed through a dynamical exchange in a
globular cluster, which would explain the strange companion, the wide
orbit, and the large space velocity, but this needs to be confirmed
with detailed numerical experiments.  The currently available radio
timing data cannot determine the orbital parameters uniquely, but
further observations and astrometric measurements of this system might
help pin down its parameters and constrain formation scenarios.

\begin{figure*}
  \plotone{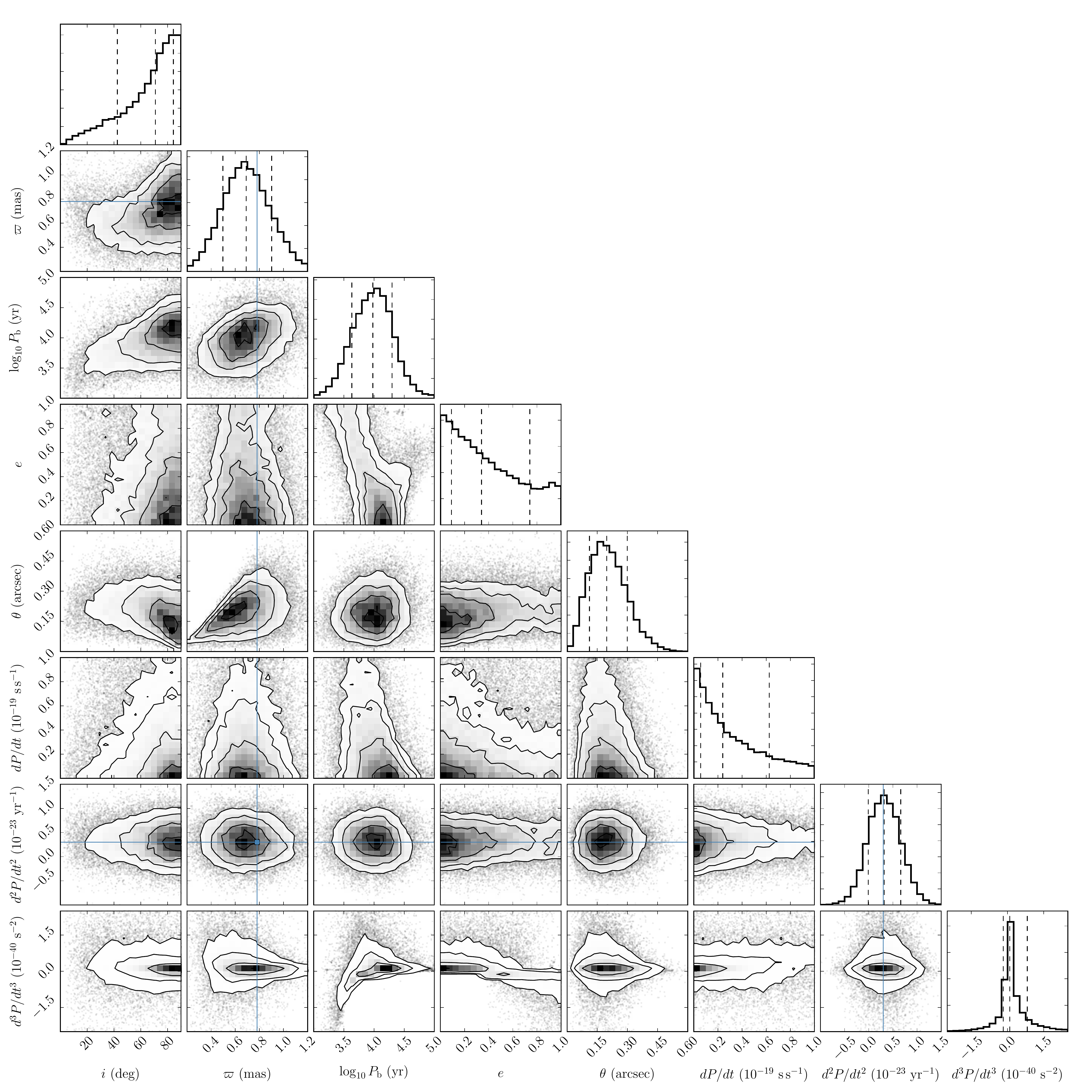}
  \caption{Markov-Chain Monte Carlo (MCMC) results for possible orbits
    for \psr.  We show the joint two-dimensional contours for the
    inclination $i$, parallax $\varpi$, binary period $P_{\rm b}$, and
    eccentricity $e$, along with derived parameters $\dot P_{\rm
      int}$, $\ddot P_{\rm orb}$, and $\dddot P_{\rm orb}$.  The
    vertical lines show the median and $\pm 1\,\sigma$ constraints on
    the one-dimensional marginal distributions, while the contours
    show 0.5, 1, 1.5, and 2-$\sigma$ joint confidence regions.  The
    blue vertical/horizontal lines are the measured value of $\varpi$
    and $\ddot P$ from \S~\ref{sec:timing}.}
  \label{fig:mcmc}
\end{figure*}

The detection of further period derivatives is one such measurement,
although care must be taken to separate dynamical period derivatives
from timing noise, dispersion measure variations, and other effects.
With observations made over a longer time span, the next-accessible
parameter of interest is period third derivative, $\dddot P$.
Positive values of $\dddot P$ are required for circular orbits and
highly favored for elliptical orbits.  As shown in
Figure~\ref{fig:mcmc}, the value should be of order $|\dddot P|\sim
1\times 10^{-40}~{\rm s}^{-2}$, or $|f_3|\sim 4\times 10^{-36}\,{\rm
  s}^{-4}$.  We estimate that such a measurement could be achieved at
$3\sigma$ significance by observations such as ours, using
dual-receiver measurements with monthly cadence, made over 15~years.
We emphasize that
dual-receiver measurements are critical: even in the existing data
set, PSR~J1024$-$0719 shows time-variable dispersion measures more
complex than a simple quadratic or cubic pattern over time, the effect
of which can only be removed through observations at widely separated
radio frequencies.

Additional progress will be made by \textit{GAIA}
\citep{2012Ap&SS.341...31D} observations of the companion to tie its
astrometry directly to the ICRF at high precision: while the distance
is unlikely to be significantly refined\footnote{ We predict a
  \textit{GAIA} magnitude $G=19.0$ \citep{2010A&A...523A..48J} for the
  companion, which leads to a parallax uncertainty of 300\,$\mu$as
  \citep{2014EAS....67...23D} and no radial velocity measurement.
  This compares with an uncertainty of 140\,$\mu$as from
  \citet{2016arXiv160105987G} or 230\,$\mu$as from
  Table~\ref{table:timing}.}, the absolute astrometry will be useful.

\acknowledgements We thank J.~Creighton, C.~Bassa, and S.~Phinney for
useful discussions.  The NANOGrav project receives support from
National Science Foundation (NSF) PIRE program award number 0968296
and NSF Physics Frontiers Center award number 1430284.  PSR's work at
NRL is supported by the Chief of Naval Research.  
Pulsar research at UBC is supported by an NSERC Discovery Grant and by
the Canadian Institute for Advanced Research.  AAM acknowledges support for this work by NASA from a Hubble Fellowship grant: HST-HF-51325.01, awarded by STScI, operated by AURA, Inc., for NASA, under contract NAS 5-26555. Part of the research was carried out at the Jet Propulsion Laboratory, California Institute of Technology, under a contract with NASA.
The National Radio Astronomy
Observatory which is a facility of the U.S. National Science
Foundation operated under cooperative agreement by Associated
Universities, Inc.

{\it Facilities:} \facility{GBT}, \facility{PO:1.5m}, \facility{Hale
  (Double Beam Spectrograph)}, \facility{VLT:Antu (FORS1)}

\bibliographystyle{apj} 

\end{document}